\documentstyle[12pt,axodraw,epsfig]{article}
\newcommand{\inv}{\left. g^0_A \right|_{\rm inv}}

\newcommand{\Frac}[2]%
{{\textstyle \frac{\mbox{\footnotesize $#1$}\rule[-0.9mm]{0mm}{1mm}}%
{\mbox{\footnotesize $#2$}\rule{0mm}{3.1mm}}}}
\renewcommand{\thefootnote}{\fnsymbol{footnote}}

\begin{document}
\begin{titlepage}
\vspace*{-12 mm}
\noindent
\begin{flushright}
\begin{tabular}{l@{}}
TUM/T39-00-09 \\
ECT-00-006 \\
hep-ph/0007293 \\
\end{tabular}
\end{flushright}
\vskip 12 mm
\begin{center}
{\Large \bf 
Axial U(1) dynamics in $\eta$ and $\eta'$ photoproduction
\footnote[1]{Work supported in part by BMBF and DFG.} } 
\\[10mm]
{\bf Steven D. Bass$^{[ab]}$, Stefan Wetzel$^{[a]}$ and Wolfram Weise$^{[a]}$}
\\[10mm]   
{\em 
$^a$ 
Physik Department, Technische Universit\"at M\"unchen, \\
D-85747 Garching, Germany } \\
\vspace{3ex}
\vspace{3ex}
{\em
$^b$
ECT*, Strada delle Tabarelle 286, I-38050 Villazzano, Trento, Italy }
\end{center}
\vskip 10 mm
\begin{abstract}
\noindent
We discuss the sensitivity of $\eta$ and $\eta'$ photoproduction near
threshold to the gluonic OZI breaking parameters in the $U_A(1)$-extended 
effective chiral Lagrangian for low-energy QCD.
Our coupled-channels analysis hints at a strong correlation between 
the gluon-induced contributions to the $\eta'$ mass and the low-energy 
$pp \rightarrow pp \eta'$ reaction and the near-threshold behaviour of
the $\gamma p \rightarrow \eta p$ cross-section. 
\end{abstract}
\end{titlepage}
\renewcommand{\labelenumi}{(\alph{enumi})}
\renewcommand{\labelenumii}{(\roman{enumii})}
\renewcommand{\thefootnote}{\arabic{footnote}}
\newpage

\section{Introduction}

Gluonic degrees of freedom play an important role in the physics of the 
flavour-singlet $J^P = 1^+$ channel \cite{okubo}
through the QCD axial anomaly \cite{zuoz}.
The most famous example is the $U_A(1)$ problem: 
the masses of the $\eta$ and $\eta'$ mesons are much greater than 
the values they would have if these mesons were pure Goldstone bosons 
associated
with spontaneously broken chiral symmetry 
\cite{weinberg,christos}.
This extra mass is induced by 
non-perturbative gluon dynamics 
\cite{thooft,hfpm,gvua1,witten} and the axial anomaly \cite{adler,bell}.
In this paper we study the effect of gluons in axial U(1) 
dynamics on $\eta$ and $\eta'$ photoproduction close to threshold.
The $\eta$ photoproduction process has been extensively studied at MAMI
\cite{mami}.
An important feature of the low-energy $\gamma p \rightarrow \eta p$ 
reaction is the prominant role of the S-wave resonance N$^*$(1535).
The $\gamma p \rightarrow \eta' p$ reaction is presently being studied 
at ELSA \cite{elsa} and Jefferson Laboratory \cite{cebaf}.
The first, and trivial, observation is that the thresholds themselves
are sensitive to OZI violation through the gluonic component to the meson 
masses. 
More challenging is to understand the role of the axial anomaly and
non-pertubative glue in the structure of the N$^*$(1535) resonance 
which couples strongly to the $\eta$, and the shape of the 
$\eta$ and $\eta'$ photoproduction cross-sections with increasing energy.
Our aim here is to study the sensitivity of the S-wave cross-sections 
to the gluonic OZI parameters in the $U_A$(1)-extended effective chiral 
Lagrangian \cite{vecca,lagran,sb99} for low-energy QCD.
We illustrate the importance of $U_A(1)$ dynamics to these processes.

The role of gluonic degrees of freedom and OZI violation in the 
$\eta'$--nucleon system has previously been investigated 
through the flavour-singlet Goldberger-Treiman relation 
\cite{venez,hatsuda} and the low-energy 
$pp \rightarrow pp \eta'$ reaction \cite{sb99}.
We frame our photoproduction discussion in the context this previous work.
The flavour-singlet Goldberger-Treiman relation connects the flavour-singlet 
axial-charge $g_A^{(0)}$ measured in polarised deep inelastic scattering 
with the $\eta'$--nucleon coupling constant $g_{\eta' NN}$.
Working in the chiral limit it reads
\begin{equation}
M g_A^{(0)} = \sqrt{3 \over 2} F_0 \biggl( g_{\eta' NN} - g_{QNN} \biggr) 
\end{equation}
where
$g_{\eta' NN}$ is the $\eta'$--nucleon coupling constant and $g_{QNN}$ 
is an OZI violating coupling which measures the one particle 
irreducible coupling of the topological charge density 
$Q = {\alpha_s \over 4 \pi} G {\tilde G}$ to the nucleon.
In Eq.(1) $M$ is the nucleon mass and $F_0$ ($\sim 0.1$GeV) renormalises 
\cite{sb99} the flavour-singlet decay constant.
The coupling constant $g_{QNN}$ is, in part, related \cite{venez} 
to the amount of spin carried by polarised gluons in a polarised proton.
The large mass of the $\eta'$ and the small value of $g_A^{(0)}$ 
\begin{equation}
\left. g^{(0)}_A \right|_{\rm pDIS} = 0.2 - 0.35
\end{equation}
extracted from deep inelastic scattering \cite{bass99,windmolders,reya}
(about a $50\%$ OZI suppression) 
point to substantial violations of the OZI rule in the flavour-singlet 
$J^P=1^+$ channel.
A large positive $g_{QNN} \sim 2.45$
is one possible explanation of the small value of $g_A^{(0)}|_{\rm pDIS}$.

\newpage

Working with the $U_A(1)$--extended chiral Lagrangian for low-energy QCD 
one finds a gluon-induced contact interaction in the 
$pp \rightarrow pp \eta'$ reaction close to threshold \cite{sb99}:
\begin{equation}
{\cal L}_{\rm contact} =
         - {i \over F_0^2} \ g_{QNN} \ {\tilde m}_{\eta_0}^2 \
           {\cal C} \
           \eta_0 \ 
           \biggl( {\bar p} \gamma_5 p \biggr)  \  \biggl( {\bar p} p \biggr)
\end{equation}
Here ${\tilde m}_{\eta_0}$ is the gluonic contribution to the mass of the
singlet 0$^-$ boson and ${\cal C}$ is a second OZI violating coupling 
which also features in $\eta'N$ scattering.
The physical interpretation of the contact term (3) 
is a ``short distance'' ($\sim 0.2$fm) interaction 
where glue is excited in the interaction region of
the proton-proton collision and 
then evolves to become an $\eta'$ in the final state.
This gluonic 
contribution to the cross-section for $pp \rightarrow pp \eta'$ 
is extra to the contributions associated with meson exchange 
models
\cite{holinde,wilkin,faldt}.
There is no reason, a priori, to expect it to be small.
Following the earlier work at SATURNE \cite{saturne} there 
is presently a vigorous experimental programme to 
investigate $\eta$ and $\eta'$ production near threshold 
in $pN$ collisions at CELSIUS \cite{celsius} and COSY \cite{cosy}.
Theoretical and experimental studies of $g_A^{(0)}$ and the $\eta'$--nucleon 
system may offer new insight into the role of gluons in chiral dynamics.

In this paper we investigate the S-wave contribution to $\eta$ and $\eta'$ 
photoproduction. The theoretical tools are the meson-baryon ``potentials'' 
derived from the low-energy $SU(3)_L \otimes SU(3)_R \otimes U_A(1)$ chiral 
effective Lagrangian together with the Lippmann-Schwinger equation.
We extend the coupled channels analysis of Kaiser, Waas and Weise 
\cite{weise} to include $\eta$--$\eta'$ mixing 
plus the coupling of axial $U_A(1)$ degrees of freedom to the nucleon.
The SU(3) coupled channels approach has been shown \cite{weise,wwa,pwave} 
to dynamically generate S-wave nucleon resonance contributions to low-energy 
hadron scattering as quasi-bound meson-baryon states.

We find that a positive value of the gluonic coupling ${\cal C}$ 
in Eq.(3) generates an attractive contribution to the 
$\eta$--nucleon and $\eta'$--nucleon 
potentials in the Lippmann-Schwinger equation.
It may, in part, contribute to the sharp rise 
in the $\eta$ photoproduction cross-section very close to threshold.
The $\gamma p \rightarrow \eta' p$ cross-section rises with increasing 
positive ${\cal C}$ and is suppressed for negative ${\cal C}$. 
Our calculations suggest that in the gedanken world where OZI 
is preserved the $S_{11}(1535)$ contribution to 
$\sigma (\gamma p \rightarrow \eta p)$ would split into 
two resonance-like structures as we reduce the gluonic contribution 
to the $\eta$ (and $\eta'$) mass.
The heavier structure is associated with a $K \Sigma$ quasi-bound state; 
the second, close to threshold, involves strong coupling to the $K \Lambda$ 
and $\eta N$ channels through the $K \Lambda \leftrightarrow \eta N$ potential.

The structure of the paper is as follows. 
In Section 2 we outline the coupled channels calculation. 
In Section 3 we briefly review the low-energy effective Lagrangian for 
$\eta$--nucleon and $\eta'$--nucleon interactions with emphasis on the 
sources of possible OZI violation.   
In Section 4 we present the numerical results. 
Finally, in Section 5, we make our conclusions.

\newpage

\section{Coupled channels calculation}

The Lippmann-Schwinger equation for the T-matrix connecting in- and
out- going channels $j$ and $i$ is
\begin{equation}
T_{ij} = V_{ij} + \sum_n {2 \over \pi} \int_0^{\infty} {\rm d}l
{l^2 \over k_n^2 -l^2 +i0} 
\biggl( {\alpha_n^2 + k_n^2 \over \alpha_n^2 + l^2} \biggr)^2  
V_{in} T_{nj}
\end{equation}
This is illustrated in Fig.1. 
We sum over two-particle intermediate states labelled by an index $n$ 
which runs from 1 to 7 and refers to
\begin{equation}
| n \rangle = 
|\pi N \rangle  ^{(1/2)} , 
|\eta N \rangle ^{(1/2)} , 
|K \Lambda \rangle ^{(1/2)} , 
|K \Sigma  \rangle ^{(1/2)} , 
|N \eta' \rangle ^{(1/2)} ,
|\pi N \rangle ^{(3/2)} , 
|K \Sigma \rangle ^{(3/2)} , 
\end{equation}
where the superscript labels isospin.
These states are connected through the energy-dependent driving terms
(which we call ``potentials'' for convenience) \cite{weise}:
\begin{equation}
V_{ij} = { \sqrt{M_i M_j} \over 4 \pi^2 F^2_{\pi} \sqrt{s} } C_{ij} ,
\end{equation}
where
$C_{ij}$
are the relative coupling strengths 
(up to a factor $- F_{\pi}^{-2}$ where $F_{\pi}$ is the pion decay
 constant).
These $C_{ij}$ for S-wave amplitudes are calculated 
from the $U_A(1)$-extended low-energy chiral Lagrangian 
-- see Eqs.(19), (23) and (25) below -- 
up to $O(p^2)$ in the meson momentum.
Working to $O(p^2)$ means at most quadratic in the meson centre of
mass energy $E_i = {s - M_i^2 + m_i^2 \over \sqrt{s}}$ where $\sqrt{s}$ 
is the total centre of mass energy, 
and $M_i$ and $m_i$ are the baryon and meson masses in channel $i$.
The potential $V_{ij}$ is iterated to all 
orders using the Lippmann-Schwinger equation (4).
In (4) $l$ denotes the relative momentum of the off-shell meson-baryon 
pair in the intermediate state $n$ and $k_n = \sqrt{E_n^2 - m_n^2}$ is 
the on-shell relative momentum.
The form-factor 
$\biggl( {\alpha_n^2 + k_n^2 \over \alpha_n^2 + l^2} \biggr)^2$
renders the $l$-integral in (4) convergent.
Here the $\alpha_n$ denotes a finite-range parameter for each channel
$n$ which has to fit to experimental data.
One expects $\alpha_n$ to lie in the range 0.5 -- 1 GeV for the SU(3) 
sector. 
The Lippmann-Schwinger equation (4) for the multi-channel T-matrix $T_{ij}$ 
is solved by simple matrix inversion
\begin{equation}
T = \biggl( 1 - V \cdot G \biggr)^{-1} V
\end{equation}
where
$G$ is the diagonal matrix
\begin{equation}
G_n 
= {2 \over \pi} \int_0^{\infty} {\rm d}l
  {l^2 \over k_n^2 -l^2 +i0} 
   \biggl( {\alpha_n^2 + k_n^2 \over \alpha_n^2 + l^2} \biggr)^2  
= {k_n^2 \over 2 \alpha_n} - {\alpha_n \over 2} - i k_n
\end{equation}
The resulting multi-channel S-matrix is 
$
S_{ij} = \delta_{ij} - 2 i \sqrt{k_i k_j} T_{ij}
$
with the total S-wave cross-section for the process $j \rightarrow i$,
\begin{equation}
\sigma_{ij} = 4 \pi {k_i \over k_j} | T_{ij} |^2 .
\end{equation}
This approach has been used to successfully describe a variety of 
meson-baryon and photoproduction processes in \cite{weise,wwa,pwave}.

\begin{figure}[h]
\epsfig{figure=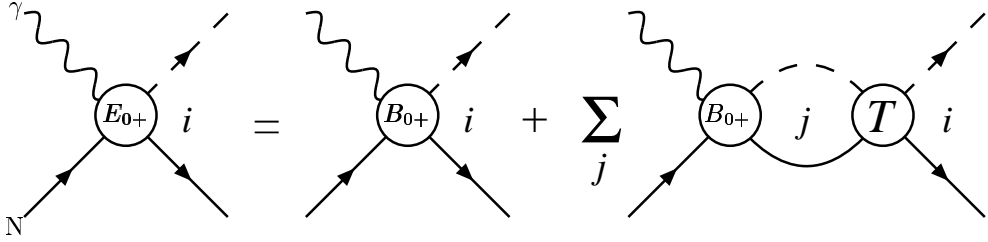, width=10cm, angle=0} 
\caption
{\it
Graphical representation of the Lippmann-Schwinger equation for S-wave 
meson photoproduction. The full, broken and wavy lines
represent baryons, mesons and the photon, respectively. 
}
\end{figure}

\vspace{1.0cm}

\section{The low-energy effective Lagrangian}

We work with the low-energy effective Lagrangian derived in 
\cite{vecca,lagran,sb99}.
Here we outline the features which are essential for our coupled channels 
calculation focussing on the key sources of possible OZI violation.
Throughout, we shall work to second order in the meson fields and the meson 
momentum $p_{\mu}$.

\subsection{Glue and the $\eta$ and $\eta'$ masses}

Starting in the meson sector, the $U_A(1)$-extended low-energy 
effective Lagrangian \cite{vecca,lagran} is
\begin{eqnarray}
{\cal L}_{\rm m} = & &
{F_{\pi}^2 \over 4} 
{\rm Tr}(\partial^{\mu}U \partial_{\mu}U^{\dagger}) 
+
{F_{\pi}^2 \over 4} {\rm Tr} \biggl[ \chi_0 \ ( U + U^{\dagger} ) \biggr]
\\ \nonumber
&+& 
{1 \over 2} i Q {\rm Tr} \biggl[ \log U - \log U^{\dagger} \biggr]
+ {3 \over {\tilde m}_{\eta_0}^2 F_{0}^2} Q^2 .
\end{eqnarray}
Here
\begin{equation}
U = \exp \ \biggl(  i {\phi \over F_{\pi}}  
                  + i \sqrt{2 \over 3} {\eta_0 \over F_0} \biggr) 
\end{equation}
is the unitary meson matrix where $\phi = \sum_k \phi_k \lambda_k$
with $\phi_k$ 
denotes the octet of would-be Goldstone bosons 
$(\pi, K, \eta_8)$
associated with 
spontaneous chiral 
$SU(3)_L \otimes SU(3)_R$ breaking, $\eta_0$ 
is the singlet boson, and $\lambda_k$ are the Gell-Mann matrices;
$\chi_0 = 
{\rm diag} [ m_{\pi}^2, m_{\pi}^2, (2 m_K^2 - m_{\pi}^2 ) ]$
is the meson mass matrix.
The pion decay constant $F_{\pi} = 92.4$MeV; 
$F_0$ renormalises the flavour-singlet decay constant.

The $U_A(1)$ gluonic potential involving the topological charge density 
$Q = {\alpha_s \over 4 \pi} G_{\mu \nu} {\tilde G}^{\mu \nu}$
generates the gluonic contribution to the $\eta$ and $\eta'$ masses.
The gluonic term $Q$ is treated as a background field with no kinetic 
term.
It may be eliminated through its equation of motion
\begin{equation}
{1 \over 2} i Q {\rm Tr} \biggl[ \log U - \log U^{\dagger} \biggr]
+ {3 \over {\tilde m}_{\eta_0}^2 F_{0}^2} Q^2 
\
\mapsto \
- {1 \over 2} {\tilde m}_{\eta_0}^2 \eta_0^2 
\end{equation}
in Eq.(10) making the mass term clear.
The $U_A(1)$ potential is also constructed to reproduce the axial anomaly 
\cite{adler,bell} in the divergence of the gauge-invariantly renormalised 
flavour-singlet axial-vector current in the effective theory,
viz.
\begin{equation}
\partial^\mu J_{\mu5} =
\sum_{k=1}^{f} 2 i \biggl[ m_k \bar{q}_k \gamma_5 q_k
\biggr]_{GI}
+ N_f
\biggl[ {\alpha_s \over 4 \pi} G_{\mu \nu} {\tilde G}^{\mu \nu}
\biggr]_{GI}^{\mu^2}
\end{equation}
where
\begin{equation}
J_{\mu 5} =
\left[ \bar{u}\gamma_\mu\gamma_5u
                  + \bar{d}\gamma_\mu\gamma_5d
                  + \bar{s}\gamma_\mu\gamma_5s \right]_{GI}^{\mu^2} .
\end{equation}
Here $N_f=3$ is the number of light flavours, the subscript
$GI$ denotes
gauge invariant renormalisation and the superscript $\mu^2$
denotes the renormalisation scale.
The Adler-Bardeen theorem \cite{bardeen} is used to constrain 
the possible $U_A(1)$ breaking terms in the effective Lagrangian.

After $Q$ is eliminated from the effective Lagrangian via (12),
we expand ${\cal L}_{\rm m}$ to ${\cal O}(p^2)$ in momentum 
keeping finite quark masses and obtain:
\begin{eqnarray}
{\cal L}_{\rm m} = & &
\sum_k {1 \over 2} \partial^{\mu} \phi_k \partial_{\mu} \phi_k 
+
{1 \over 2} \partial_{\mu} \eta_0 \partial^{\mu} \eta_0 
\ \biggl( {F_{\pi} \over F_0} \biggr)^2 \ 
- {1 \over 2} {\tilde m}_{\eta_0}^2 \eta_0^2 \ 
\\ \nonumber
&-& {1 \over 2} 
m_{\pi}^2 \biggl( 2 \pi^+ \pi^- + \pi_0^2 \biggr)
- m_K^2 \biggl( K^+ K^- + K^0 {\bar K}^0 \biggr)
- {1 \over 2} \biggl( {4 \over 3} m_K^2 - {1\over 3} m_{\pi}^2 \biggr)
  \eta_8^2
\\ \nonumber
&-& {1 \over 2} \biggl( {2 \over 3} m_K^2 + {1\over 3} m_{\pi}^2 \biggr)
    \ \biggl( {F_{\pi} \over F_0} \biggr)^2 \ \eta_0^2
    + {4 \over 3 \sqrt{2}} \biggl( m_K^2 - m_{\pi}^2 \biggr) 
    \ \biggl( {F_{\pi} \over F_0} \biggr) \eta_8 \eta_0
\end{eqnarray}
In the chiral limit, $\chi_0 = 0$, gluons contribute a finite mass 
\begin{equation}
m_{\eta_0}^2 = 
{\tilde m}_{\eta_0}^2
\biggl( {F_0 \over F_{\pi}} \biggr)^2
\end{equation}
to the singlet $\eta_0$.

For finite quark masses $\eta$--$\eta'$ mixing occurs. 
For simplicity 
we work in the one mixing angle scheme \cite{gilman}:
\begin{eqnarray}
| \eta \rangle &=& 
\cos \theta \ | \eta_8 \rangle - \sin \theta \ | \eta_0 \rangle
\\ \nonumber
| \eta' \rangle &=& 
\sin \theta \ | \eta_8 \rangle + \cos \theta \ | \eta_0 \rangle
\end{eqnarray}
The physical value of $\theta$ is taken as -18 degrees 
\cite{gilman,frere}.
The masses of the physical $\eta$ and $\eta'$ mesons are found 
by diagonalising the $(\eta_8, \eta_0)$ mass matrix which follows 
from Eq.(15).

The value of $F_0$ is usually determined from the decay rate for
$\eta' \rightarrow 2 \gamma$.
In QCD one finds the relation \cite{shore}
\begin{equation}
{2 \alpha \over \pi} = 
\sqrt{3 \over 2} F_0 \biggl( g_{\eta' \gamma \gamma} - g_{Q \gamma \gamma} 
\biggr) 
\end{equation}
The observed decay rate \cite{twogamma} is consistent \cite{gilman} with 
the OZI prediction for $g_{\eta' \gamma \gamma}$ {\it if} $F_0$ and 
$g_{Q \gamma \gamma}$ take their OZI values: 
$F_0 \simeq F_{\pi}$ and $g_{Q \gamma \gamma} = 0$.
Motivated by this observation it is common to take $F_0 \simeq F_{\pi}$.
In this paper we shall allow $F_{\pi} / F_0$ to vary between 0.8 and 1.25.

\subsection{OZI violation and the $\eta'$--baryon interaction}

The low-energy effective Lagrangian (10) is readily extended 
to include $\eta$--nucleon and $\eta'$--nucleon coupling.
Working to $O(p)$ in the meson momentum the chiral 
Lagrangian for meson-baryon coupling is 
\begin{eqnarray}
{\cal L}_{\rm mB} &=&
{\rm Tr} \ {\overline B} (i \gamma_{\mu} D^{\mu} - M_0) B
\\ \nonumber
&+& F \
{\rm Tr} \biggl( {\overline B} \gamma_{\mu} \gamma_5 [a^{\mu}, B]_{-}
  \biggr)
+ D \ 
{\rm Tr} \biggl( {\overline B} \gamma_{\mu} \gamma_5 \{a^{\mu}, B\}_{+}
  \biggr)
\\ \nonumber
&+&
{i \over 3}
K \ {\rm Tr} \biggl({\overline B} \gamma_{\mu} \gamma_5 B \biggr)
    {\rm Tr} \biggl(U^{\dagger} \partial^{\mu} U \biggr) 
- {{\cal G}_{QNN} \over 2 M_0} \partial^{\mu} Q 
  {\rm Tr} \biggl( {\overline B} \gamma_{\mu} \gamma_5 B \biggr) 
+ 
{{\cal C} \over F_0^4} Q^2 {\rm Tr} \biggl( {\overline B} B \biggr) 
\end{eqnarray}
Here
\begin{equation}
B =\
\left(\begin{array}{ccc} 
{1 \over \sqrt{2}} \Sigma^0 + {1 \over \sqrt{6}} \Lambda & \Sigma^+ & p \\
\\
\Sigma^- & -{1 \over \sqrt{2}} \Sigma^0 + {1\over \sqrt{6}} \Lambda & n \\
\\
\Xi^- & \Xi^0 & -{2 \over \sqrt{6}} \Lambda
\vphantom{\inv}  
\end{array}\right) 
\end{equation}
denotes the baryon octet and $M_0$ denotes the baryon mass in the chiral 
limit.
In Eq.(19)
\begin{equation}
D_{\mu} B 
= \partial_{\mu} B - i e {\cal A}_{\mu} [ {\rm e_q}, B ]
+ {1 \over 8 F_{\pi}^2} [  [ \phi, \partial_{\mu} \phi ], B ] + ...
\end{equation}
is the chiral covariant derivative where 
${\rm e_q} = {\rm diag}[+{2 \over 3}, -{1 \over 3}, -{1 \over 3}]$ 
is the quark charge matrix and ${\cal A}_{\mu}$ is the photon field;
$a_{\mu}$ is the axial-vector current operator
\begin{equation}
a_{\mu} = 
- {1 \over 2 F_{\pi}} \partial_{\mu} \phi 
- {1 \over 2 F_0} \sqrt{2 \over 3} \partial_{\mu} \eta_0 
+ {i e \over 2 F_{\pi}} {\cal A}_{\mu} [ {\rm e_q}, \phi ]  + ...
\end{equation}
The chiral covariant derivative in (21) is independent of the singlet
boson $\eta_0$.
The SU(3) couplings are
$F= 0.459 \pm 0.008$ and $D= 0.798 \pm 0.008$ \cite{fec}.
The $U_A(1)$ coupling $K$ is dimensionless and 
the two gluonic couplings ${\cal G}_{QNN}$ and ${\cal C}$ 
both have mass dimension -3. 
In general, one may expect OZI violation wherever a coupling involving
the $Q$-field occurs.

We shall work consistently to $O(p^2)$.
This means that we include the chiral corrections to the baryon masses:
\begin{equation}
{\cal L}^{(2)}_{\rm mass} =
b_D {\rm Tr} \biggl( {\bar B} \{ \chi_+, B \}_+ \biggr) +
b_F {\rm Tr} \biggl( {\bar B}  [ \chi_+, B  ]_- \biggr) +
b_0 {\rm Tr} \biggl( {\bar B} B \biggr) {\rm Tr}   \chi_+
\end{equation}
where 
\begin{eqnarray}
\chi_+ &=& 
\biggl( \xi^{\dagger} \chi_0 \xi^{\dagger} + \xi \chi_0 \xi \biggr)
\\ \nonumber
&=&
2 \chi_0 + 
2 \Biggl(  - {1 \over 8 F_{\pi}^2} \{ \phi \{ \phi, \chi_0 \}_+ \}_+
           - {1 \over 3 F_0^2    } \chi_0 \eta_0^2 
           - {1 \over 2} \sqrt{2 \over 3} {1 \over F_{\pi} F_0 } \eta_0 
             \{ \phi, \chi_0 \}_+ + ... \Biggr)
\end{eqnarray}
and $\xi = U^{1 \over 2}$.
We also include the heavy-baryon terms \cite{weise}
\begin{eqnarray}
{\cal L}^{(2)}_{\rm HB} = & &
  2 d_D {\rm Tr} \biggl( {\overline B} \{ (v \cdot a)^2, B \} \biggr)
+ 2 d_F {\rm Tr} \biggl( {\overline B}  [ (v \cdot a)^2, B  ] \biggr)
\\ \nonumber
&+& 2 d_0 {\rm Tr} \biggl( {\overline B} B \biggr)
        {\rm Tr} \biggl( (v \cdot a)^2 \biggr)
  + 2 d_1 {\rm Tr} \biggl( {\overline B} v \cdot a \biggr)
        {\rm Tr} \biggl( v \cdot a B \biggr)
\\ \nonumber
&+&
2 d_K {\rm Tr} \biggl( {\overline B} B \biggr)
 \biggl(
 v_{\mu}
  {\rm Tr} \biggl[ \partial^{\mu} U \cdot U^{\dagger} \biggr]
 \biggr)^2
\end{eqnarray}
where $v_{\mu} =  (1; {\vec 0})$ is a four-velocity.
(We refer to \cite{manohar,meissner} for reviews of heavy-baryon theory.)
Note the new $U_A(1)$ term proportional to $d_K$. 
Motivated by the lack of any kinetic term for $Q$ 
in the meson Lagrangian (10) we do not include a 
term proportional to $(v_{\mu} \partial^{\mu} Q)^2$
although the $U_A(1)$ term $d_K$ is understood to 
contain possible OZI violation.

The parameters $b_D$ and $b_F$ are determined from the baryon mass 
shifts: one finds \cite{weise}
$b_D = +0.066$GeV$^{-1}$ and $b_F = -0.213$GeV$^{-1}$.
The value of $M_0$ in Eq.(19) is fixed by the size of $b_0$ 
which is constrained by the size of the pion-nucleon
sigma term.
The five $d_i$
parameters 
are fit to scattering data as done in Refs. \cite{weise,pwave}.

When we eliminate $Q$ through its equation of motion the $Q$ dependent
terms in the effective Lagrangian become:
\begin{eqnarray}
{\cal L}_Q &=& {1 \over 12} {\tilde m}_{\eta_0}^2 \
   \biggl[ \ - 6 \eta_0^2 \  
           - \ {\sqrt{6} \over M_0} \ {\cal G}_{QNN} \ F_0 \ 
               \partial^{\mu} \eta_0 \
           {\rm Tr} \biggl( {\bar B} \gamma_{\mu} \gamma_5 B \biggr) 
\\ \nonumber
& & \ \ \ \ \ \ \ \ 
           + \ {\cal G}_{QNN}^2 \ F_0^2 \
             \biggl( {\rm Tr} {\bar B} \gamma_5 B \biggr)^2 \
           + \ 2 \ {\cal C} \ {{\tilde m}_{\eta_0}^2 \over F_0^2 } \
           \eta_0^2 \ {\rm Tr} \biggl( {\bar B} B \biggr) 
\\ \nonumber
& & \ \ \ \ \ \ \ \ 
           - \ {\sqrt{6} \over 3 M_0 F_0} \ {\cal G}_{QNN} \
           {\cal C} {\tilde m}_{\eta_0}^2 
           \eta_0 \ \partial^{\mu}  
           {\rm Tr} \biggl( {\bar B} \gamma_{\mu} \gamma_5 B \biggr)  \
           {\rm Tr} \biggl( {\bar B} B \biggr)
+ ... \biggr] 
\end{eqnarray}
The third, fourth and fifth terms in the Lagrangian (26) 
are contact terms associated with the gluonic potential in $Q$.
The last term in Eq.(26) is the gluonic contact term 
(3)
in the low-energy $pp \rightarrow pp \eta'$ reaction
with
$g_{QNN} \equiv
 \sqrt{1 \over 6} {\cal G}_{QNN} F_0 { \tilde m}^2_{\eta_0}$.
The term
\begin{equation}
{\cal L}^{(3)}_{\rm contact} =
  {1 \over 6 F_0^2} \ {\cal C} \ 
  {\tilde m}_{\eta_0}^4 \ \eta_0^2 \ {\rm Tr} \biggl( {\bar B} B \biggr) .
\end{equation}
is potentially important to $\eta$--nucleon and $\eta'$--nucleon scattering 
processes. 
It will contribute to the intermediate state in our coupled channels 
calculation as an {\it attractive} potential (6) 
in the Lippmann-Schwinger equation (4) for {\it positive} ${\cal C}$.

The OZI violating Lagrangian ${\cal L}_Q$ is proportional 
to ${\tilde m}_{\eta_0}^2$
which vanishes in the formal OZI limit.
Phenomenologically, the large masses of the $\eta$ and 
$\eta'$ mesons imply that there is no reason, a priori, 
to expect ${\cal L}_Q$ to be small.
We note that large $N_c$ predictions for the $\eta'$--nucleon system 
should be treated with care. Assuming a continuous large $N_c$ limit, 
one finds $m^2_{\eta'} \sim 1/N_c$ and $M_N \sim N_c$ 
whereas $m_{\eta'}$ is greater than $M_N$ in the real world!
The large $N_c$ approximation is badly violated in the $U_A(1)$ channel.

Some hint on the possible size of the $U_A(1)$ parameters in the chiral 
Lagrangian comes from the flavour-singlet Goldberger-Treiman relation (1).
If OZI were exact in the singlet $1^+$ channel
the Ellis-Jaffe sum-rule would hold and one 
would find $g_A^{(0)} = g_A^{(8)} \simeq 0.6$.
If one attributes the OZI suppression of the flavour-singlet 
axial-charge 
extracted from polarised deep inelastic scattering,
$\left. g^{(0)}_A \right|_{\rm pDIS} = 0.2 - 0.35$, 
to 
the gluonic coupling $g_{QNN}$ then one finds
$g_{QNN} \sim 2.45$
or
${\cal G}_{QNN} \simeq +60$GeV$^{-3}$.
If one further takes this value and saturates the COSY measurement 
\cite{cosy} of 
the low-energy $pp \rightarrow pp \eta'$ cross-section with the contact 
term (3) then one finds $|{\cal C}| \simeq 2$GeV$^{-3}$ \cite{sb99}.
Of course, in practice, other processes will contribute to the measured 
$pp \rightarrow pp \eta'$ cross-section.
However, this simple estimate does provide a handle on 
the possible size of the OZI violation parametrised by ${\cal C}$.

\section{Results}

To investigate the effect of OZI violation in the flavour-singlet $J^P=1^+$ 
channel on the $\eta$ and $\eta'$ photoproduction processes we work with 
the Lippmann-Schwinger equation (4). The potentials for SU(3) 
(sub-)processes are listed in \cite{weise}.
We have generalised these potentials to include $\eta$--$\eta'$ mixing and 
the coupling of $U_A(1)$ degrees of freedom to the nucleon.
The results are listed in the Appendix.

Our aim here is to investigate the qualitative effect of different 
possible OZI violations on the $\eta$ and $\eta'$ photoproduction 
cross-sections 
rather than to make quantitative predictions for $\eta'$ scattering and 
production processes.
We look for definite general effects in the cross-sections as we vary 
the OZI parameters.
\footnote{
Several caveats are in order if one wishes to extend the analysis 
presented here from a qualitative to a quantitative description 
and to make numerical predictions for $\eta'$ production using 
this type of approach -- 
two of them associated with the large $\eta'$ mass. 
First, the potentials in \cite{weise} are derived assuming 
$E_i \ll M_0$
for meson energies, 
which may provide a reasonable approximation for the $\eta$ but 
not for physical $\eta'$ photoproduction.
For the physical $\eta'$ mass
we are close to the limit of valid application of the effective 
theory, $\mu \sim 4 \pi F_{\pi}$.}
We isolate which channels are the most important and check 
the consistency of our results by observing that they also
hold when we decrease each of the meson masses by a uniform scaling
factor $\lambda \simeq 0.75$ with the $b_i$ and $d_i$ held fixed.

The potentials have the following general features.
Starting with $F_0=F_{\pi}$ and ${\cal C} = d_K =0$ 
the $U_A(1)$ potentials $C_{\eta \eta}$ and $C_{\eta' \eta'}$ 
are repulsive, and $C_{\eta \eta'}$ is attractive.
For ${\cal C} = d_K =0$ clearly the most prominent effect 
in these three potentials comes from the heavy-baryon terms (25);
they contribute
$+1.1 E_{\eta}^2$ in $C_{\eta \eta}$,
$-0.4 E_{\eta} E_{\eta'}$ in $C_{\eta \eta'}$, 
and 
$+1.1 E_{\eta'}^2$ in $C_{\eta' \eta'}$.
The Born term contributions to the meson-baryon 
scattering potentials turn out to be very small.
One finds a strong attractive coupling, $-1.1 E_K E_{\eta'}$, 
of 
the $|\eta' N \rangle$ to $|K \Sigma \rangle^{(1/2)}$ states.
Each of these $|\eta N \rangle$, $|\eta' N \rangle$ and 
$|K \Sigma \rangle^{(1/2)}$
intermediate states play an important role in generating
the $\gamma p \rightarrow \eta' p$ cross-section.

In Fig 2 we show a fit to the $\eta$ photoproduction cross-section 
from our S-wave calculation with $\eta$-$\eta'$ mixing included and 
the OZI violating couplings and $d_K$ turned off.
We also show the S-wave contribution to the $\eta'$ photoproduction 
cross-section which is generated with the same set of $b_0$ and $d_i$ 
parameters.
All figures are calculated with $\alpha_{\eta' N}=1.5$GeV. 
The S-wave indeed dominates $\eta$ photoproduction close to threshold
through the $S_{11}(1535)$.
This is not so for $\eta'$ photoproduction where S-waves are expected
to account only for part of the $\eta'$ cross-section even close to
threshold.
The general results presented below do not depend strongly on
$\alpha_{\eta' N}$, for $\alpha_{\eta' N}$ between 1 and 2GeV.
Figs. 2-6 are calculated with the flavour-singlet coupling 
$g_{\eta_0 NN}$ set equal to its OZI value, 
corresponding to $X=0.58$ in Eq.(30) below.
Approximately, the $\eta'$ photoproduction cross-section 
grows proportional to the singlet coupling $g_{\eta_0 NN}$ and 
the $\eta$ photoproduction cross-section is independent of $g_{\eta_0 NN}$

\vspace{0.8cm}

\begin{figure}[h]
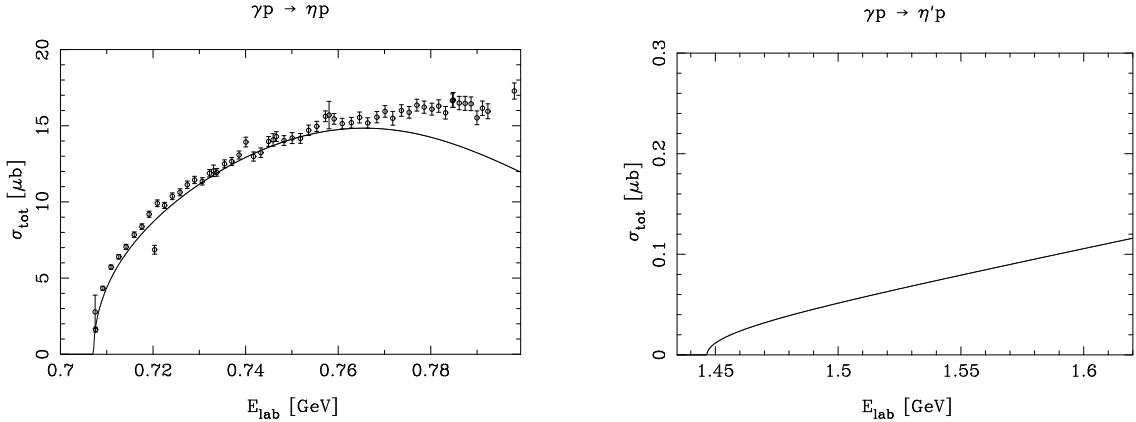

\epsfig{figure=2a1.ps, width=5.5cm, angle=-90} \hfill
\epsfig{figure=2a2.ps, width=5.5cm, angle=-90}
\caption{
\it
Fit to the low-energy $\eta$ photoproduction data \cite{mami}
with ${\cal C} = d_K =0$, 
and the $\eta'$ photoproduction cross-section generated 
from the same fit parameters}
\end{figure}

For positive ${\cal C}$ the gluonic term (27) is attractive in each 
of the 
$C_{\eta \eta}$, $C_{\eta \eta'}$ and $C_{\eta' \eta'}$ potentials.
Motivated by the COSY measurement of 
$pp \rightarrow pp \eta'$ 
close
to threshold we vary ${\cal C}$ from -4 to +4GeV$^{-3}$, 
which we consider a reasonable guess at its maximum likely magnitude.
With increasing
positive ${\cal C}$ in our ``reasonable range'' 
the $\eta$ and $\eta'$ cross-sections are enhanced very 
close to threshold and the broad maximum in the $\eta$ production
cross-section shifts closer to threshold.
With increasing negative ${\cal C}$ 
they are suppressed
and the peak in the $\eta$ production
cross-section shifted to higher energy.
We show this in Fig.3 keeping the $b_0$ and $d_i$ parameters fixed at 
their OZI values for ${\cal C} = 0,\pm 2$. 
It is not unreasonable that a large gluonic production mechanism in the 
low-energy  $pp \rightarrow pp \eta'$ reaction and the shape of the 
$S_{11}(1535)$ resonance may be correlated.

\vspace{0.8cm}

\begin{figure}[h]
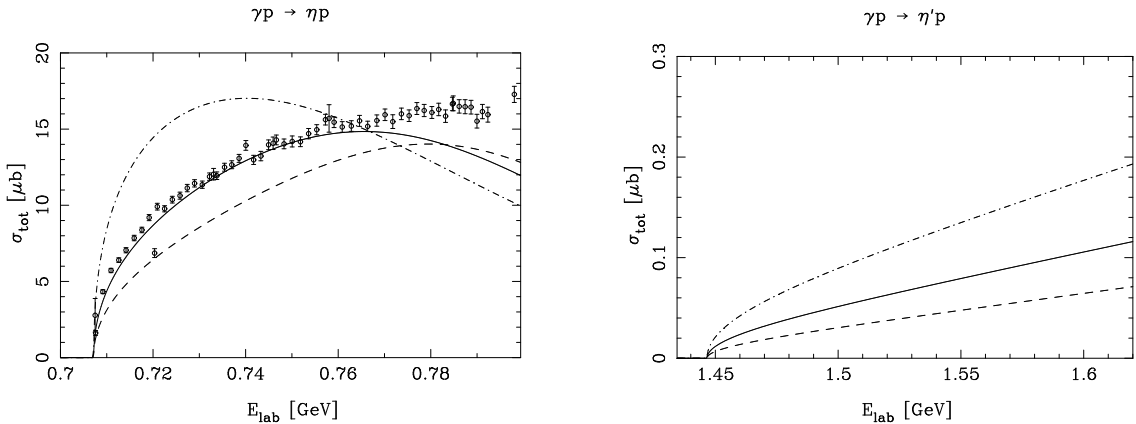

\epsfig{figure=2b1.ps, width=5.5cm, angle=-90} \hfill
\epsfig{figure=2b2.ps, width=5.5cm, angle=-90}
\caption{
\it
The fit from Fig. (2), solid line, together 
with the $\eta$ and $\eta'$ photoproduction
cross-sections produced by varying the gluonic
coupling ${\cal C} =+2$, dot-dashed line, and 
${\cal C} =-2$, dashed line, with all other parameters held fixed.
}
\end{figure}

\clearpage

\begin{figure}[h]
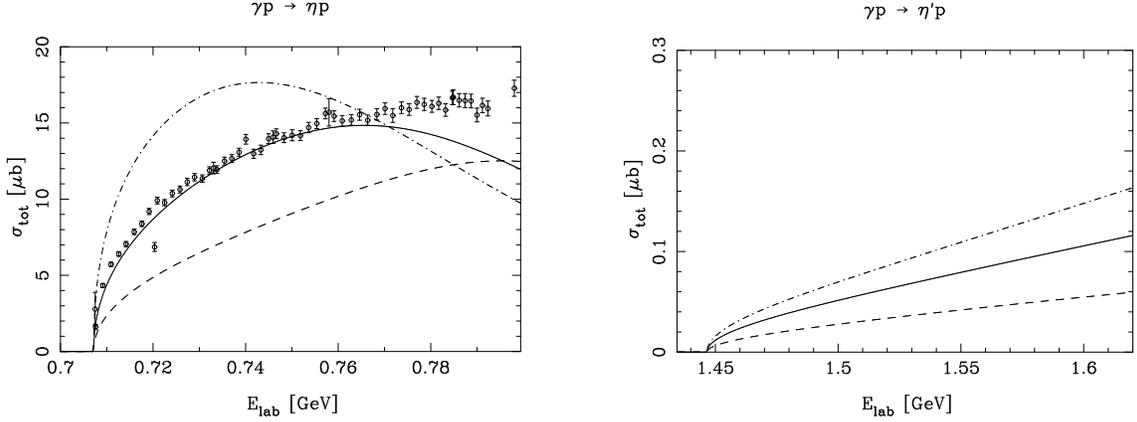

\epsfig{figure=2c1.ps, width=5.5cm, angle=-90} \hfill
\epsfig{figure=2c2.ps, width=5.5cm, angle=-90}
\caption{
\it
The fit from Fig. (2), solid line, together 
with the $\eta$ and $\eta'$ photoproduction
cross-sections produced by varying 
${F_{\pi} \over F_0} =0.8$ 
, dot-dashed line, and 
${F_{\pi} \over F_0} =1.25$, 
dashed line, with all other parameters held fixed.
}
\end{figure}

Varying $F_{\pi}/F_0$ between 0.8 and 1.25 we find that the photoproduction 
cross-sections are enhanced close to threshold for larger values of $F_0$
-- see Fig. 4, corresponding to an 
enhanced heavy-baryon repulsion in $C_{\eta' \eta'}$.

We also consider the effect of the heavy-baryon term proportional to
$d_K$ in Eq.(25).
We find that the $\eta'$ photoproduction cross-section is enhanced
for values of $d_K$ in a small window around -0.035. We show this in Fig.5.

\begin{figure}[h]
\begin{center}
\epsfig{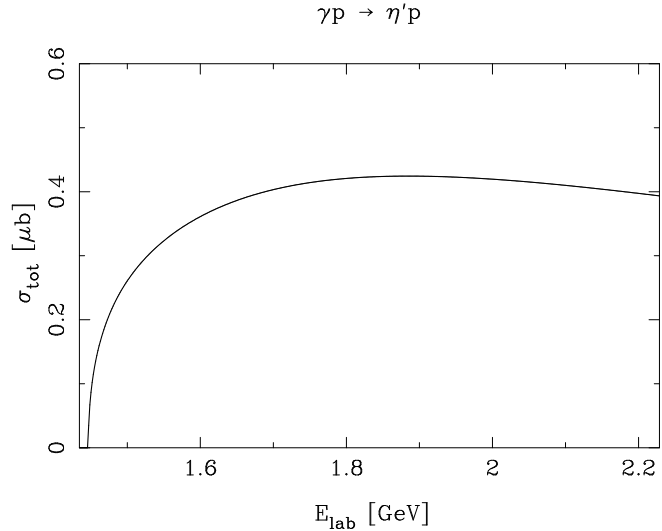}
\end{center}
\caption{
\it
The $\eta'$ photoproduction cross-section is enhanced with a small 
negative $d_K = -0.035$. Here ${\cal C} = 0$ and $F_0 = F_{\pi}$.}
\end{figure}

Finally, we consider the effect of varying the gluonic contribution 
to the singlet mass:
${\tilde m}^2_{\eta_0}$.
Varying the ${\tilde m}^2_{\eta_0}$ means varying the masses of 
both the $\eta$ and $\eta'$ mesons which, in turn, means varying 
the $\eta$--$\eta'$ mixing angle $\theta$.
Setting $F_0=F_{\pi}$ one may diagonalise the $\eta$--$\eta'$ 
mass matrix 
which follows from (15) 
to obtain the masses of the physical $\eta$ and $\eta'$ mesons:
\begin{equation}
m^2_{\eta', \eta} = (m_{\rm K}^2 + {\tilde m}_{\eta_0}^2 /2) 
\pm {1 \over 2} 
\sqrt{(2 m_{\rm K}^2 - 2 m_{\pi}^2 - {1 \over 3} {\tilde m}_{\eta_0}^2)^2 
   + {8 \over 9} {\tilde m}_{\eta_0}^4} .
\end{equation}
If we turn off the gluon mixing term, 
then one finds
$m_{\eta'} = \sqrt{2 m_{\rm K}^2 - m_{\pi}^2}$ 
and
$m_{\eta} = m_{\pi}$.
Summing over the two eigenvalues yields \cite{vecca} 
\begin{equation}
m_{\eta}^2 + m_{\eta'}^2 = 2 m_K^2 + {\tilde m}_{\eta_0}^2 .
\end{equation}
Substituting the physical values of 
$(m_{\eta}^2 + m_{\eta'}^2)$ in Eq.(29) and 
$m_K^2$ yields ${\tilde m}_{\eta_0}^2 = 0.73$GeV$^2$,
which corresponds to
$m_{\eta} = 499$MeV and $m_{\eta'} = 984$MeV.
The value ${\tilde m}_{\eta_0}^2 = 0.73$GeV$^2$ 
corresponds to an $\eta - \eta'$ 
mixing angle $\theta \simeq - 18$ degrees -- the physical value.
As we decrease ${\tilde m}^2_{\eta_0}$ the $S_{11}(1535)$ 
peak in the $\eta$ photoproduction cross-section splits into 
two resonance-like structures -- Fig.6.
The heavier structure is associated with a $K \Sigma$ quasi-bound state; 
the second, close to threshold,
involves strong coupling to the 
$K \Lambda$ and $\eta N$ channels through 
the $K \Lambda \leftrightarrow \eta N$ potential.

\begin{figure}[h]
\begin{center}
\epsfig{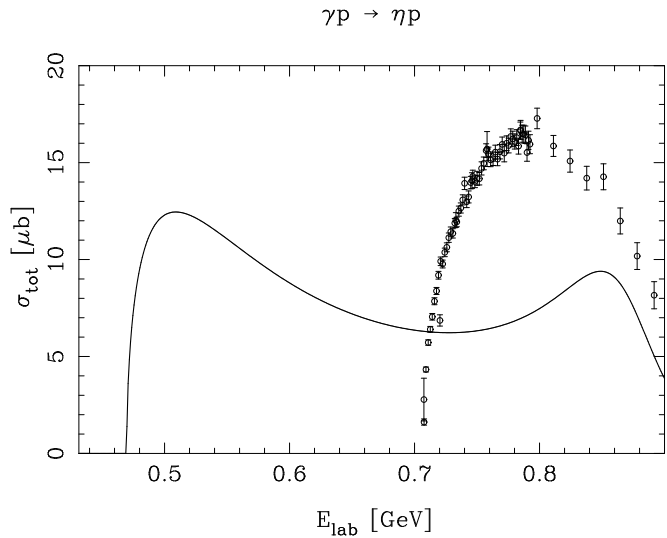} 
\end{center}
\caption{
\it
Reducing the gluonic contribution to the $\eta$ and $\eta'$
masses by changing
${\tilde m}^2_{\eta_0}$ 
from 0.73 to 0.25 GeV$^2$, the $\eta$ and $\eta'$ masses in (28) 
become 0.15 GeV and 0.59 GeV respectively
and the $S_{11}(1535)$ splits into two resonance-like structures.
}
\end{figure}

\section{Conclusions}

Gluon dynamics through OZI violation play a potentially important 
role in the $\eta$ and $\eta'$ nucleon interactions.
Through the axial anomaly gluonic degrees of freedom 
generate 
about 300 -- 400 MeV of the $\eta$ and $\eta'$ masses,
and thereby affect the threshold behaviour of $\eta$ and $\eta'$
photo- and proton-induced production.
They also induce new gluonic coupling constants in the $\eta$ and
$\eta'$ nucleon interactions.
Our coupled channels analysis hints at a strong correlation 
between the gluonic contribution to the low-energy 
$pp \rightarrow pp \eta'$ reaction and the sharp rise of the 
$\gamma p \rightarrow \eta p$ cross-section close to threshold.
It follows that gluonic effects in the axial U(1) channel may be important 
in the structure of the N$^*$(1535) resonance which couples strongly to the 
$\eta$.
The S-wave contribution to the $\eta'$ photoproduction cross-section 
close to threshold also increases with increasing positive gluonic coupling 
${\cal C}$.
For $\eta'$ photoproduction higher partial waves are likely 
to be important in building up the total $\eta'$ photoproduction cross-section.

\vspace{0.5cm}

{\bf Acknowledgements} \\

SDB thanks for the European Centre ECT*, Trento for a Senior Visiting
Fellowship where part of this work was completed.
We thank B. Borasoy and N. Kaiser for helpful discussions.
We also thank J. Link and B. Ritchie for communication about present and 
forthcoming experiments.

\newpage

\section*{Appendix: Potentials for the $U_A(1)$ sector}

\subsection*{The S-wave potential for photoproduction}

To simplify the notation we define
\begin{equation}
X = 
\biggl( 2D + 2K + {\cal G}_{QNN} F_0^2 {{\tilde m}_{\eta_0}^2 \over 2M_0}
 \biggr)
\end{equation}

The S-wave potentials for photoproduction become
\begin{equation}
B_{0+}^{(2)} =
{e M_N \over 8 \pi F_{\pi} \sqrt{3 s}} 
\ (3F-D) \
\biggl[
\cos \theta \ Y_{\eta} + \sin \theta \ Y_{\eta'} \biggr]
\end{equation}
and
\begin{equation}
B_{0+}^{(5)} =
{e M_N \over 8 \pi F_{\pi} \sqrt{3 s}} {F_{\pi} \over F_0} 
\ X \
\biggl[
\cos \theta \ Y_{\eta'} - \sin \theta \ Y_{\eta} \biggr]
\end{equation}
where
\begin{equation}
Y_{\phi} = - {1 \over 3M_0} 
\biggl( 2E_{\phi} + { m_{\phi}^2 \over E_{\phi} } \biggr)
\end{equation}
In Eqs.(31) and (32) $Y_{\eta}$ corresponds to $\eta$ 
photoproduction and $Y_{\eta'}$ corresponds to $\eta'$ 
photoproduction.

\subsection*{S-wave potentials for meson-baryon scattering}

Following \cite{weise} we define
\begin{eqnarray}
S_{ab} &=& {E_a E_b \over 2 M_0} \\ \nonumber
U_{ab} &=& {1 \over 3 M_0} 
\biggl( 2 m_a^2 + 2 m_b^2 + 
        {m_a^2 m_b^2 \over E_a E_b} - {7 \over 2} E_a E_b
\biggr)
\\
\end{eqnarray}

\subsection*{Potentials with $\eta$ and $\eta'$}

Working with the notation of Eq.(5) one finds:
\begin{eqnarray}
C_{1 \eta} &=& 
{1 \over 4} \ (F+D) \
\biggl( \cos \theta \ (3F-D) \ - \ 
        \sqrt{2} \ \sin \theta \ X \ {F_{\pi} \over F_0} \biggr)
\biggl( S_{\pi \eta} + U_{\pi \eta} \biggr) 
\\ \nonumber
& & 
- \ ( d_D + d_F ) \
\biggl( \cos \theta - \sqrt{2} \ \sin \theta \ {F_{\pi} \over F_0} \biggr) 
\ E_{\pi} E_{\eta} 
\\ \nonumber
& & 
+ \ m_{\pi}^2 \ ( b_D + b_F ) \
\biggl[ 2 \cos \theta \ - \ 2 \sqrt{2} \ \sin \theta 
        {F_{\pi} \over F_0}   \biggr]
\end{eqnarray}

\begin{eqnarray}
C_{1 \eta'} &=& 
{1 \over 4} \ (F+D) \
\biggl( \sin \theta \ (3F-D) \ + \ 
        \sqrt{2} \ \cos \theta \ X \ {F_{\pi} \over F_0} \biggr)
\biggl( S_{\pi \eta'} + U_{\pi \eta'} \biggr) 
\\ \nonumber
& & 
- \ ( d_D + d_F ) \
\biggl( \sin \theta + \sqrt{2} \ \cos \theta \ {F_{\pi} \over F_0} \biggr) 
\ E_{\pi} E_{\eta'} 
\\ \nonumber
& & 
+ \ m_{\pi}^2 \ ( b_D + b_F ) \
\biggl[ 2 \sin \theta \ + \ 2 \sqrt{2} \ \cos \theta \
        {F_{\pi} \over F_0}   \biggr]
\end{eqnarray}

\begin{eqnarray}
C_{\eta 3} &=& \
\Biggl[ ({1 \over 12} D^2 - {3 \over 4} F^2) \ \cos \theta 
\ + \ {\sqrt{2} \over 12} (3F+D) \sin \theta \ X \ {F_{\pi} \over F_0} 
\Biggr]
\ S_{K \eta} 
\\ \nonumber
& &
+ \ \Biggl[ {1 \over 2} \ D (F + {1 \over 3} D) \ \cos \theta 
\ + \ {\sqrt{2} \over 12} (3F+D) \sin \theta \ X \ {F_{\pi} \over F_0} 
\Biggr]
\ U_{K \eta} 
\\ \nonumber
& & 
- \
\Biggl[ ({1 \over 6} d_D + {1 \over 2} d_F + d_1) \ \cos \theta \ 
      + {\sqrt{2} \over 3} ( d_D + 3 d_F ) \ \sin \theta \ {F_{\pi} \over F_0} 
\Biggr]
\ E_K E_{\eta} 
\\ \nonumber
& & 
+ \
(b_D + 3 b_F) \
\Biggl[
({5 \over 6} m_K^2 - {1 \over 2} m_{\pi}^2) \ \cos \theta \ 
  + {2 \over 3} \sqrt{2} \ m_K^2 \ \sin \theta \ {F_{\pi} \over F_0} \Biggr]
\\ \nonumber
& &
+ {3 \over 8} ( E_K + \cos \theta \ E_{\eta} )
+ {3 \over 16 M_0} 
\biggl[ E_K^2 - m_K^2 + \cos \theta \ (E_{\eta}^2 - m_{\eta}^2)  \biggr]
\end{eqnarray}

\begin{eqnarray}
C_{\eta' 3} &=& \
\Biggl[ ({1 \over 12} D^2 - {3 \over 4} F^2) \ \sin \theta 
\ - \ {\sqrt{2} \over 12} (3F+D) \cos \theta \ X \ {F_{\pi} \over F_0} 
\Biggr]
\ S_{K \eta} 
\\ \nonumber
& &
+ \ \Biggl[ {1 \over 2} \ D (F + {1 \over 3} D) \ \sin \theta 
\ - \ {\sqrt{2} \over 12} (3F+D) \cos \theta \ X \ {F_{\pi} \over F_0} 
\Biggr]
\ U_{K \eta} 
\\ \nonumber
& & 
- \
\Biggl[ ({1 \over 6} d_D + {1 \over 2} d_F + d_1) \ \sin \theta \ 
      - {\sqrt{2} \over 3} ( d_D + 3 d_F ) \ \cos \theta \ {F_{\pi} \over F_0} 
\Biggr]
\ E_K E_{\eta'} 
\\ \nonumber
& & 
+ \
(b_D + 3 b_F) \
\Biggl[
({5 \over 6} m_K^2 - {1 \over 2} m_{\pi}^2) \ \sin \theta \ 
  - {2 \over 3} \sqrt{2} m_K^2 \cos \theta {F_{\pi} \over F_0} \Biggr]
\\ \nonumber
& &
+ {3 \over 8} ( E_K + \sin \theta \ E_{\eta'} )
+ {3 \over 16 M_0} 
\biggl[ E_K^2 - m_K^2 + \sin \theta \ (E_{\eta'}^2 - m_{\eta'}^2)  \biggr]
\end{eqnarray}

\begin{eqnarray}
C_{\eta 4} &=& 
\Biggl[
{1 \over 4} (3F-D) (D-F) \ \cos \theta \ - \ 
            {1 \over 2 \sqrt{2}} \ (D-F) \ \sin \theta \ X \ 
            {F_{\pi} \over F_0}  \Biggr]    
\ S_{K \eta} 
\\ \nonumber
& &
+ \
\Biggl[
{1 \over 2} D (D-F) \ \cos \theta \ - 
            {1 \over 2 \sqrt{2}} \ (D-F) \ \sin \theta \ X \ 
            {F_{\pi} \over F_0}  \Biggr]    
\ U_{K \eta} 
\\ \nonumber
& &
+ \ 
(d_D - d_F) \
\Biggl[ {1 \over 2} \ \cos \theta \ + \ \sqrt{2} \ \sin \theta \
        {F_{\pi} \over F_0}  \Biggr] 
\ E_K E_{\eta} 
\\ \nonumber
& & 
\ + \ (b_D - b_F) \
\Biggl[
( {3 \over 2} m_{\pi}^2 -{5 \over 2} m_K^2 ) \ \cos \theta \ 
-
2 \sqrt{2} \ m_K^2  \ \sin \theta \ {F_{\pi} \over F_0} \Biggr]
\\ \nonumber
& &
\ + {3 \over 8} (E_K \ + \ \cos \theta \ E_{\eta} ) 
  + {3 \over 16 M_0} 
    \Biggl[ E_K^2 - m_K^2 + \cos \theta \ (E_{\eta}^2 - m_{\eta}^2) \Biggr]
\end{eqnarray}

\begin{eqnarray}
C_{\eta' 4} &=& 
\Biggl[
{1 \over 4} (3F-D) (D-F) \ \sin \theta \ + \ 
            {1 \over 2 \sqrt{2}} \ (D-F) \ \cos \theta \ X \ 
            {F_{\pi} \over F_0}  \Biggr]    
\ S_{K \eta} 
\\ \nonumber
& &
+ \
\Biggl[
{1 \over 2} D (D-F) \ \sin \theta \ + 
            {1 \over 2 \sqrt{2}} \ (D-F) \ \cos \theta \ X \ 
            {F_{\pi} \over F_0}  \Biggr]    
\ U_{K \eta} 
\\ \nonumber
& &
+ \ 
(d_D - d_F) \
\Biggl[ {1 \over 2} \ \sin \theta \ - \ \sqrt{2} \ \cos \theta \
        {F_{\pi} \over F_0}  \Biggr] 
\ E_K E_{\eta'} 
\\ \nonumber
& & 
\ + \ (b_D - b_F) \
\Biggl[
( {3 \over 2} m_{\pi}^2 -{5 \over 2} m_K^2 ) \ \sin \theta \ 
+
2 \sqrt{2} \ m_K^2  \ \cos \theta \ {F_{\pi} \over F_0} \Biggr]
\\ \nonumber
& &
\ + {3 \over 8} (E_K \ + \ \sin \theta \ E_{\eta'} ) 
  + {3 \over 16 M_0} 
    \Biggl[ E_K^2 - m_K^2 + \sin \theta \ (E_{\eta'}^2 - m_{\eta'}^2) \Biggr]
\end{eqnarray}

\begin{eqnarray}
C_{\eta \eta} &=& 
\Biggl[
{1 \over 12} \ (3F-D)^2 \ \cos^2 \theta 
-
{\sqrt{2} \over 24} \ (3F-D) \ X \ {F_{\pi} \over F_0} \ \sin 2 \theta \
\\ \nonumber
& &
\ \ \ \ \ \ \ \ \ \ 
+
{1 \over 6} \ X^2 \ \biggl( {F_{\pi} \over F_0} \biggr)^2 \ \sin^2 \theta
\Biggr] 
\ \biggl( S_{\eta \eta} + U_{\eta \eta} \biggr) 
\\ \nonumber
& &
\ - {1 \over 3} {\cal C} {\tilde m}^4 
    \biggl( {F_{\pi} \over F_0} \biggr)^2
    \sin^2 \theta \ 
\\ \nonumber
& &
\ +
\Biggl[
\biggl(d_F - {5 \over 3} d_D - 2 d_0 \biggr) \ \cos^2 \theta \
- \ {\sqrt{2} \over 2} \
  \biggl( {d_D \over 3} - d_F \biggr) {F_{\pi} \over F_0} \ \sin 2 \theta
\\ \nonumber
& &
\ \ \ \ \ \ \ \ \ \ +
\biggl( - 2 d_0 - {4 \over 3} d_D + 24 d_K \biggr)
\biggl( {F_{\pi} \over F_0} \biggr)^2 \ \sin^2 \theta
\Biggr]  
\ E_{\eta} E_{\eta}
\\ \nonumber
& &
\ +
\biggl[
{16 \over 3} m_K^2 (b_D - b_F + b_0) 
+ 2 m_{\pi}^2 ({5 \over 3}b_F - b_D - {2 \over 3}b_0)
\biggr]
\cos^2 \theta \ 
\\ \nonumber
& &
\ \ \ \ \ \ \ \ \ \ 
- {1 \over 2}
\biggl[
{\sqrt{2} \over 2} m_{\pi}^2 ( 3 b_D - b_F + 4 b_0 ) 
    +   2 \sqrt{2} m_K^2     ( - b_D + b_F -   b_0 ) 
\biggr]
\ {F_{\pi} \over F_0} \ \sin 2 \theta
\\ \nonumber
& & 
\ \ \ \ \ \ \ \ \ \ +
{4 \over 3}
\biggl[  m_{\pi}^2 ( 2b_F + b_0) 
       + 2 m_K^2 (b_D - b_F + b_0)
\biggr]
\ \biggl( {F_{\pi} \over F_0} \biggr)^2 \ \sin^2 \theta \ 
\end{eqnarray}


\begin{eqnarray}
C_{\eta \eta'} &=& 
\Biggl[
{1 \over 24} \ (3F-D)^2 \ \sin 2 \theta 
-
{\sqrt{2} \over 12} \ (3F-D) \ X \ {F_{\pi} \over F_0} \ \cos 2 \theta \
\\ \nonumber
& &
\ \ \ \ \ \ \ \ \ \ 
-
{1 \over 12} \ X^2 \ \biggl( {F_{\pi} \over F_0} \biggr)^2 \ \sin 2 \theta
\Biggr]
\
\biggl( S_{\eta \eta'} + U_{\eta \eta'} \biggr) 
\\ \nonumber
& &
\ + {1 \over 6} \ {\cal C} \ {\tilde m}^4 \
    \biggl( {F_{\pi} \over F_0} \biggr)^2 \
    \sin 2 \theta \ 
\\ \nonumber
& &
\ +
\Biggl[ {1 \over 2}
\biggl(d_F - {5 \over 3} d_D - 2 d_0 \biggr) \ \sin 2 \theta \
- \ \sqrt{2} \
  \biggl( {d_D \over 3} - d_F \biggr) {F_{\pi} \over F_0} \ \cos 2 \theta
\\ \nonumber
& &
\ \ \ \ \ \ \ \ \ \ 
- {1 \over 2} \
\biggl( - 2 d_0 - {4 \over 3} d_D + 24 d_K \biggr) \
\biggl( {F_{\pi} \over F_0} \biggr)^2 \ \sin 2 \theta
\Biggr]  
\ E_{\eta} E_{\eta'}
\\ \nonumber
& &
\ +
{1 \over 2}
\biggl[
{16 \over 3} m_K^2 (b_D - b_F + b_0) 
+ 2 m_{\pi}^2 ({5 \over 3}b_F - b_D - {2 \over 3}b_0)
\biggr]
\sin 2 \theta \ 
\\ \nonumber
& &
\ \ \ \ \ \ \ \ \ \ 
- \
\biggl[
{\sqrt{2} \over 2} m_{\pi}^2 ( 3 b_D - b_F + 4 b_0 ) 
    +   2 \sqrt{2} m_K^2     ( - b_D + b_F -   b_0 ) 
\biggr]
\ {F_{\pi} \over F_0} \ \cos 2 \theta
\\ \nonumber
& & 
\ \ \ \ \ \ \ \ \ \ 
- 
{2 \over 3}
\biggl[  m_{\pi}^2 ( 2b_F + b_0) 
       + 2 m_K^2 (b_D - b_F + b_0)
\biggr]
\ \biggl( {F_{\pi} \over F_0} \biggr)^2 \ \sin 2 \theta \ 
\end{eqnarray}

\begin{eqnarray}
C_{\eta' \eta'} &=& 
\Biggl[
{1 \over 12} \ (3F-D)^2 \ \sin^2 \theta 
+
{\sqrt{2} \over 24} \ (3F-D) \ X \ {F_{\pi} \over F_0} \ \sin 2 \theta \
\\ \nonumber
& & 
\ \ \ \ \ \ \ \ \ \ 
+
{1 \over 6} \ X^2 \ \biggl( {F_{\pi} \over F_0} \biggr)^2 \ \cos^2 \theta
\Biggr]
\ \biggl( S_{\eta' \eta'} + U_{\eta' \eta'} \biggr) 
\\ \nonumber
& &
\ - \ {1 \over 3} \ {\cal C} \ {\tilde m}^4 
    \biggl( {F_{\pi} \over F_0} \biggr)^2 \
    \cos^2 \theta \ 
\\ \nonumber
& &
\ +
\Biggl[
\biggl(d_F - {5 \over 3} d_D - 2 d_0 \biggr) \ \sin^2 \theta \
+ \ {\sqrt{2} \over 2} \
  \biggl( {d_D \over 3} - d_F \biggr) {F_{\pi} \over F_0} \ \sin 2 \theta
\\ \nonumber
& &
\ \ \ \ \ \ \ \ \ \ 
+
\biggl( - 2 d_0 - {4 \over 3} d_D + 24 d_K \biggr)
\biggl( {F_{\pi} \over F_0} \biggr)^2 \ \cos^2 \theta
\Biggr]  
\ E_{\eta'} E_{\eta'}
\\ \nonumber
& &
\ +
\biggl[
{16 \over 3} m_K^2 (b_D - b_F + b_0) 
+ 2 m_{\pi}^2 ({5 \over 3}b_F - b_D - {2 \over 3}b_0)
\biggr]
\ \sin^2 \theta \ 
\\ \nonumber
& &
\ \ \ \ \ \ \ \ \ \ 
+ {1 \over 2}
\biggl[
{\sqrt{2} \over 2} m_{\pi}^2 ( 3 b_D - b_F + 4 b_0 ) 
    +   2 \sqrt{2} m_K^2     ( - b_D + b_F -   b_0 ) 
\biggr]
\ {F_{\pi} \over F_0} \ \sin 2 \theta
\\ \nonumber
& & 
\ \ \ \ \ \ \ \ \ \ +
{4 \over 3}
\biggl[  m_{\pi}^2 ( 2b_F + b_0) 
       + 2 m_K^2 (b_D - b_F + b_0)
\biggr]
\ \biggl( {F_{\pi} \over F_0} \biggr)^2 \ \cos^2 \theta \ 
\end{eqnarray}

\end{document}